\documentclass[10pt, conference, letterpaper]{IEEEtran}
\IEEEoverridecommandlockouts
 \usepackage{arydshln}  
\usepackage{multirow}

\usepackage{booktabs}
\usepackage{array}
\usepackage{float}

\usepackage{cite}
\usepackage{amsmath,amssymb,amsfonts}
\usepackage{algorithmic}
\usepackage{graphicx}
\usepackage{textcomp}
\usepackage{makecell}
\usepackage{siunitx}
\usepackage{booktabs}
\usepackage[table,dvipsnames]{xcolor}
\usepackage{subfigure}
\usepackage[ruled,linesnumbered, vlined]{algorithm2e}
\usepackage{balance}
\usepackage{enumitem}
\usepackage{multirow} 
\usepackage{ulem}
\usepackage{amssymb}
\usepackage{bbding}
 
\usepackage{textcomp}
\usepackage{makecell}
\usepackage{pifont}
\usepackage{tikz}

\newcommand*\circledwhite[1]{\tikz[baseline=(char.base)]{
            \node[shape=circle,draw,black, fill=white, minimum size=10pt, inner sep=1pt] (char) {#1};}}

\usepackage{bbding}
\usepackage{pifont}
\usepackage{tikz}
\def\BibTeX{{\rm B\kern-.05em{\sc i\kern-.025em b}\kern-.08em
    T\kern-.1667em\lower.7ex\hbox{E}\kern-.125emX}}

\usepackage{url}
\definecolor{rowcolor}{HTML}{E2F7FE}

\begin{document}

\title{\textit{Venus}: An Efficient Edge Memory-and-Retrieval System for VLM-based Online Video Understanding}

\author{
    \IEEEauthorblockN{Shengyuan Ye$^\blacklozenge$$^*$, Bei Ouyang$^\blacklozenge$$^*$, Tianyi Qian$^\blacklozenge$, Liekang Zeng$^\blacktriangle$, Mu Yuan$^\blacktriangle$,
    }
    \IEEEauthorblockN{Xiaowen Chu$^\vartriangle$, Weijie Hong$^\lozenge$, Xu Chen$^{\blacklozenge\dagger}$}
    \IEEEauthorblockA{$^\blacklozenge$School of Computer Science and Engineering, Sun Yat-sen University, Guangzhou, China}
    \IEEEauthorblockA{$^\blacktriangle$Department of Information Engineering, The Chinese University of Hong Kong, Hong Kong SAR, China}
    \IEEEauthorblockA{$^\vartriangle$Data Science and Analytics Thrust, HKUST (Guangzhou), Guangzhou, China}
    \IEEEauthorblockA{$^\lozenge$Shenzhen Smart City Communications Co., Ltd., China}
    \IEEEauthorblockA{\{yeshy8, ouyb9, qianty\}@mail2.sysu.edu.cn, \{lkzeng, muyuan\}@cuhk.edu.hk}
    \IEEEauthorblockA{xwchu@ust.hk, hongweijie@smartcitysz.com, chenxu35@mail.sysu.edu.cn}
    \thanks{$*$: Equal contributions. $\dagger$: Corresponding author.}
}

\maketitle

\begin{abstract}
Vision-language models (VLMs) have demonstrated impressive multimodal comprehension capabilities and are being deployed in an increasing number of online video understanding applications.
While recent efforts extensively explore advancing VLMs' reasoning power in these cases, deployment constraints are overlooked, leading to overwhelming system overhead in real-world deployments.
To address that, we propose \texttt{Venus}, an on-device memory-and-retrieval system for efficient online video understanding. 
\texttt{Venus} proposes an edge–cloud disaggregated architecture that sinks memory construction and keyframe retrieval from cloud to edge, operating in two stages.
In the \textit{ingestion stage}, \texttt{Venus} continuously processes streaming edge videos via scene segmentation and clustering, where the selected keyframes are embedded with a multimodal embedding model to build a hierarchical memory for efficient storage and retrieval.
In the \textit{querying stage}, \texttt{Venus} indexes incoming queries from memory, and employs a threshold-based progressive sampling algorithm for keyframe selection that enhances diversity and adaptively balances system cost and reasoning accuracy.
Our extensive evaluation shows that \texttt{Venus} achieves a $15\times$–$131\times$ speedup in total response latency compared to state-of-the-art methods, enabling real-time responses within seconds while maintaining comparable or even superior reasoning accuracy.
\end{abstract}

\section{Introduction}
Building on the advancements of large language models, vision-language models (VLMs) \cite{liu2023visual, zhang2024video} have been pre-trained on vast collections of image-text pairs, allowing them to learn rich representations that capture fine-grained visual-linguistic correlations. These capabilities have driven significant progress across various visual understanding tasks \cite{tang2025adaptive, feng2025vision}.
With these capabilities, an increasing number of commercial and research organizations are self-hosting VLMs on public or private cloud infrastructures and exposing them via standardized APIs \cite{openai_api}. This growing accessibility to advanced vision-language reasoning has empowered developers and researchers to integrate VLMs into real-world systems, fueling the rapid development of \textit{edge intelligence applications} that advance visual understanding, including multimodal personal assistants \cite{li2024personal, ye2025jupiter}, smart surveillance \cite{fan2025ai}, and city scene reasoning \cite{sun20243d}.

\begin{figure}[t!]
    \setlength{\abovecaptionskip}{-0.1cm}
    \centering
    \includegraphics[width=0.95\linewidth]{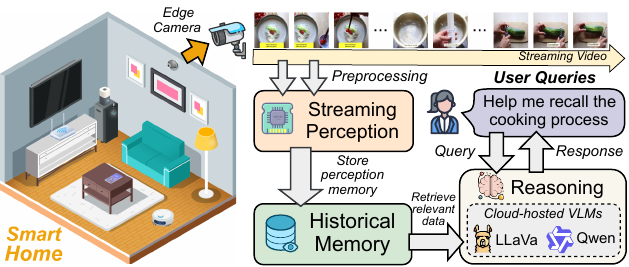}
    \caption{An online video understanding application empowered by VLMs.}
    \label{fig:intro}
    \vspace{-15pt}
\end{figure}

Edge cameras are ubiquitous, spanning a wide range of public and private infrastructures.
Online video from edge cameras is continuously recorded and retrieved on demand, making intelligent analysis and understanding of streaming edge video a longstanding focus in edge AI research \cite{hsieh2018focus, xu2020approximate}.
Prior to the VLM era, CNNs had long served as the backbone of online video understanding.
While lightweight, CNNs are insufficient for high-level semantic retrieval and cross-modal reasoning, driving a paradigm shift toward VLM-based workflows that incorporate rich visual-linguistic reasoning \cite{niu2025ovo, huang2025online}.
Figure \ref{fig:intro} abstracts a VLM-powered online video understanding application deployed in a smart-home scenario, where family members can initiate queries targeting current or historical segments of recorded video at any time.
The application consists of three key modules: (1) \textit{Streaming Perception Module}: Edge cameras equipped with onboard compute continuously capture video of household scenes and process multimodal data on the fly.
(2) \textit{Historical Memory Module}: Historical video streams are constructed and stored in a memory database to support efficient and accurate retrieval of query-relevant information.
(3) \textit{Reasoning Module}: Upon receiving a user query, cloud-hosted VLMs are leveraged to perform reasoning and generate answers, with historical memory providing contextual grounding.

Emerging research efforts in the AI community \cite{liu2025bolt, li2024llava,luo2024video} have explored the effectiveness of VLMs for video understanding, but most efforts focus on algorithmic performance while overlooking deployment-related constraints, limiting their direct applicability in edge intelligence scenarios.
Fully cloud-based deployment leverages powerful server-side computation to reduce on-device processing, but requires uploading the entire relevant video, incurring significant communication latency.
Some methods \cite{liu2025bolt, tang2025adaptive} improve reasoning accuracy by performing frame selection prior to inference. An alternative deployment strategy executes the selection algorithm on the edge, allowing only the selected frames to be uploaded to the cloud. While this significantly reduces communication overhead, the limited compute capacity of edge devices incurs substantial on-device processing costs.
Beyond overwhelmed communication and computation overhead, existing methods are inherently restricted to offline execution and are unsuitable for online streaming applications that involve online user queries and require long-term contextual memory.

To address these limitations, we propose \texttt{Venus}, a novel on-device memory-and-retrieval system tailored for efficient online edge video understanding.
\texttt{Venus} is characterized by two essential capabilities:
(1) \textit{Memory}: Enable real-time video perception by embedding multimodal inputs and inserting vector representations into a historical memory, supporting efficient recall and low-latency query reasoning.
(2) \textit{Retrieval}: Support natural language queries with precise and adaptive retrieval of keyframes from memory, minimizing data transmission to cloud-hosted VLMs and reducing communication and computation overhead for low-latency responses.
Based on this neat principle, \texttt{Venus} leverages a novel edge–cloud disaggregated architecture that performs multimodal memory construction and keyframe retrieval on the edge, with VLM reasoning in the cloud. The architecture comprises two main stages:
In the \textit{ingestion stage}, the edge camera continuously records streaming video. To support real-time on-device perception, \texttt{Venus} applies scene segmentation and frame clustering to eliminate redundancy, selecting cluster centroids as index frames to represent visually similar content. Each index frame is then embedded into a vector representation using a multimodal embedding model (MEM), supplemented by lightweight auxiliary models. The resulting indexed vectors, together with the original frames in their associated clusters, form a hierarchical memory that supports accurate and efficient recall and retrieval.
The \textit{querying stage} is activated upon receiving a query, which is encoded using the same MEM and computed similarity scores with vectors in memory. Rather than relying on a greedy Top-K algorithm, \texttt{Venus} designs a sampling-based strategy to balance relevance and diversity in keyframe selection. To overcome the limitations of a fixed sampling budget, \texttt{Venus} introduces a threshold-driven progressive sampling algorithm that adaptively adjusts the number of selected frames, enabling a flexible trade-off between system cost and reasoning accuracy.
We implement the \texttt{Venus} system and conduct a performance evaluation on realistic edge testbeds. 
Our extensive evaluation shows that \texttt{Venus} achieves a $15\times$–$131\times$ speedup in total response latency compared to state-of-the-art methods, while maintaining comparable or even superior reasoning accuracy.

The main contributions are summarized as follows:
\begin{itemize}[leftmargin=*]
\item We revisit deployment limitations in existing algorithms in AI community and propose an edge–cloud disaggregated architecture to serve VLM-based online video understanding.
\item We design a scene segmentation and clustering module to enable real-time on-device perception. The streaming video frames are embedded and organized into a hierarchical memory to support accurate and efficient recall and retrieval.
\item We propose a threshold-driven progressive sampling algorithm for keyframes selection that enhances diversity and adaptively balances system cost and reasoning accuracy.
\item We incorporate and implement our designs into a system \texttt{Venus} and evaluate it on realistic edge devices. Our extensive evaluation shows that \texttt{Venus} achieves a $15\times$–$131\times$ speedup in total response latency compared to baselines.
\end{itemize}

\section{Background and Motivation}

\begin{figure}[t]
    \setlength{\abovecaptionskip}{-0.1cm}
    \centering
    \includegraphics[width=0.92\linewidth]{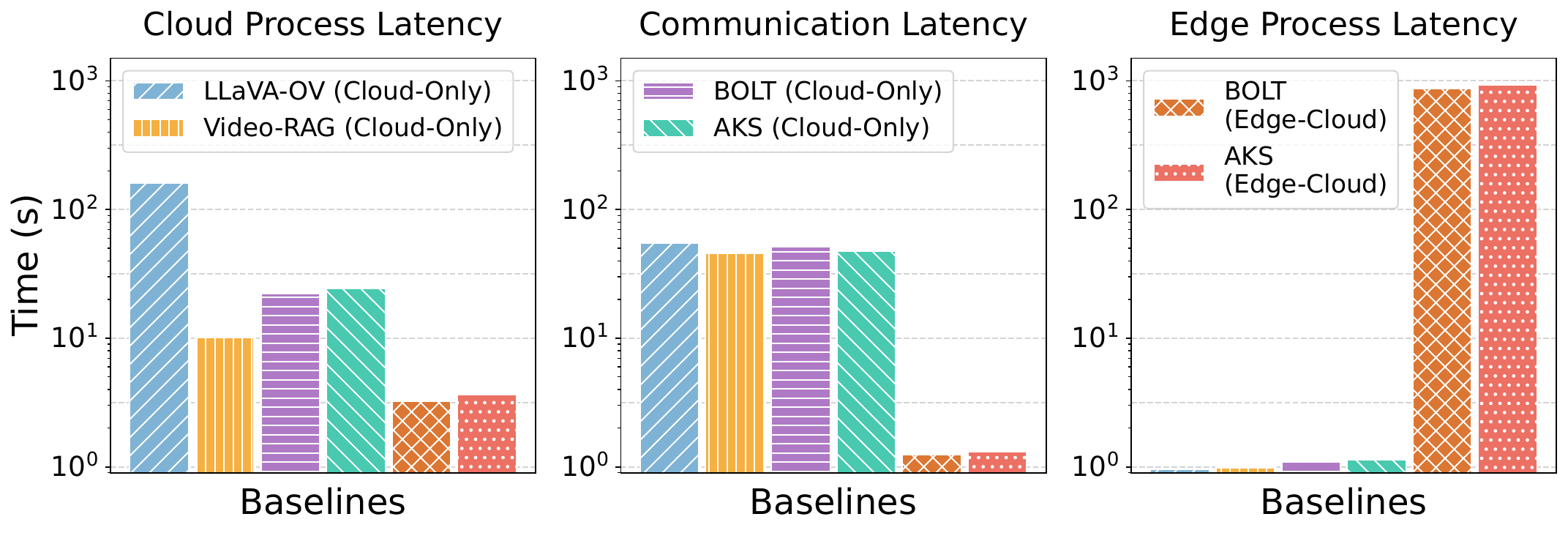}
    \caption{Latency breakdown for video understanding task, including communication, cloud, and on-device processing. Measured on NVIDIA AGX Orin using an 8 FPS EgoSchema video. VLMs run on a server with one NVIDIA L40S GPU. Video-RAG, BOLT, and AKS sample 32 frames.}
    \label{fig:motivation_breakdown}
    \vspace{-15pt}
\end{figure}

\subsection{VLM-Based Online Video Understanding Applications}
VLMs demonstrate strong capabilities in multimodal perception \cite{liu2023visual, zhang2024video} and are increasingly integrated into edge applications to enhance visual understanding. 
A VLM-based online video understanding application employs an edge camera for continuous video capture, enabling natural language queries over current or historical video segments at any time, with real-time, accurate responses to users powered by a VLM.
For example, queries may involve recalling recent activities such as the cooking process or verifying whether an elderly person took their medication in the afternoon.

\textbf{Difference with Traditional Video Analysis Workflow.}
Prior to the VLM era, CNNs had long served as the backbone of traditional online video analysis systems. Numerous on-device \cite{li2020reducto, hsieh2018focus} or edge–cloud collaborative \cite{wang2024gecko, zhang2021towards,yan2024visflow} studies have proposed specialized optimizations for CNN-based real-time video analytics pipelines.
However, fundamental architectural and functional disparities between CNNs and VLMs hinder the transferability of optimizations from CNN-based to VLM-based pipelines. 
Also, VLMs remain impractical for edge deployment due to their substantially higher resource demands than CNNs, and many state-of-the-art VLMs are still closed-source, making them hard to self-host. These factors result in VLM capabilities being typically accessible only via black-box APIs hosted by cloud providers.
\textit{These differences motivate the redesign of workflows for the VLM era.}

\subsection{Limitations of Existing VLM-based Video Understanding}
\label{sec:motivation-exp}
Recent advances in the AI community \cite{li2024llava, zhang2023video} have introduced VLMs into video understanding tasks to enhance semantic representation and cross-modal integration.
Existing VLM-based video understanding methods generally overlook deployment-related constraints, which prevent their direct applicability in edge intelligence scenarios.

LLaVA-OneVision \cite{li2024llava} processes every video frame using 196 tokens without any reduction, resulting in approximately 6 million tokens for a one-hour video at 8 FPS. This approach introduces significant video transmission and inference overhead.
Video-RAG \cite{luo2024video} and MDF \cite{han2024self} apply uniform sampling or redundant frame removal to reduce video token redundancy. Although efficient, these query-agnostic strategies assume equal frame importance and often underperform in complex real-world scenarios. Sampling too sparsely risks missing critical content, while oversampling introduces redundancy that increases transmission and inference costs.

Other recent methods such as BOLT \cite{liu2025bolt} and AKS \cite{tang2025adaptive} leverage vision-language similarity to identify frames relevant to the input query. They typically use contrastive language-image pretraining encoders \cite{radford2021learning} to associate frames with textual descriptions, followed by tailored selection algorithms that only retrieve query-relevant frames for downstream video understanding with VLMs.
Their adaptive selection helps balance informativeness and inference efficiency in video understanding.
However, these methods still require transferring the entire relevant video to the cloud for processing, making significant communication latency unavoidable.
An alternative is to execute these algorithms on edge devices, allowing only the selected frames to be uploaded to the cloud, which can significantly reduce transmission overhead. 
However, such a frame-wise algorithm design is impractical for online scenarios, as edge devices often lack the computational capacity to run transformer-based encoders on every frame in real time.

\begin{figure}[t!]
    \setlength{\abovecaptionskip}{-0.1cm}
    \centering
    \includegraphics[width=0.95\linewidth]{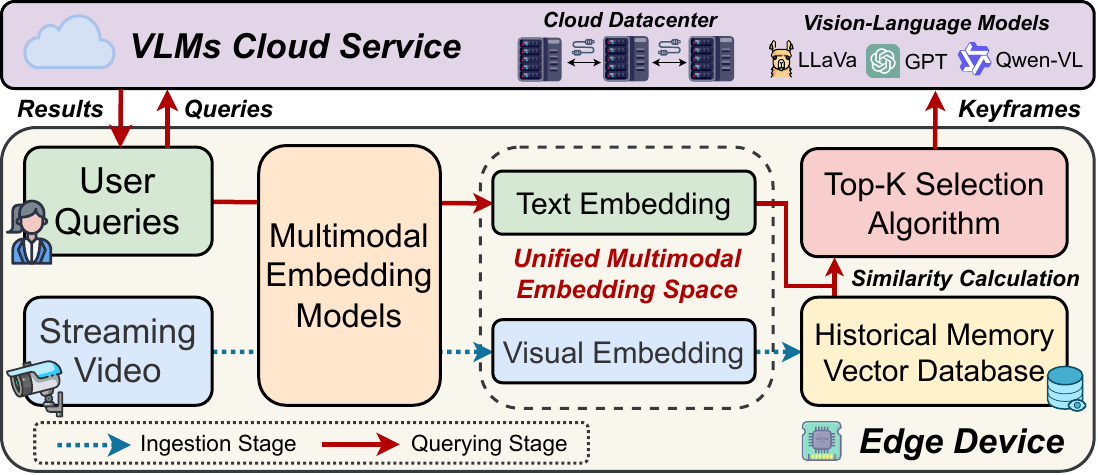}
    \caption{An edge-cloud disaggregated architecture.}
    \label{fig:strawman}
    \vspace{-15pt}
\end{figure}

Figure \ref{fig:motivation_breakdown} compares the latency breakdown of existing methods under different deployment strategies. Fully cloud-based approaches (\textit{Cloud-Only}) leverage powerful server-side computation to reduce on-device processing overhead, but require uploading the entire video stream to the cloud, which incurs communication latency that accounts for up to $80\%$ of the total response latency. This overhead grows substantially with longer video durations.
An alternative deployment strategy executes the frame selection algorithms of BOLT and AKR on the edge, allowing only the selected frames to be uploaded to the cloud (\textit{Edge-Cloud}). While this significantly reduces communication overhead, the limited compute capacity of edge devices leads to up to $924$s of on-device latency due to frame-wise processing.
\textit{Beyond overwhelmed communication and computation overhead, all aforementioned methods are inherently limited to offline execution and are unsuitable for online streaming scenarios that demand online querying and the ability to memorize long-term context.}

\subsection{Our Design Goals}
\label{sec:design-goal}
As summarized earlier, none of the existing methods fully meet the requirements for direct deployment in our online video understanding applications. 
Alternatively, we revisit these limitations and endeavor to propose an efficient memory-and-retrieval system tailored for online edge video understanding, guided by the following design goals:
(1) Support real-time video processing with continuous injection of multimodal representations into a historical memory database, enabling efficient recall and low-latency query-based reasoning.
(2) Support user-friendly querying through natural language input, and enable accurate and adaptive retrieval of key frames and multimodal context from memory to minimize the number of data sent to cloud-hosted VLMs, thereby reducing communication and computation overhead for low-latency responses.

\begin{figure}[t]
    \setlength{\abovecaptionskip}{-0.1cm}
    \centering
    \includegraphics[width=0.85\linewidth]{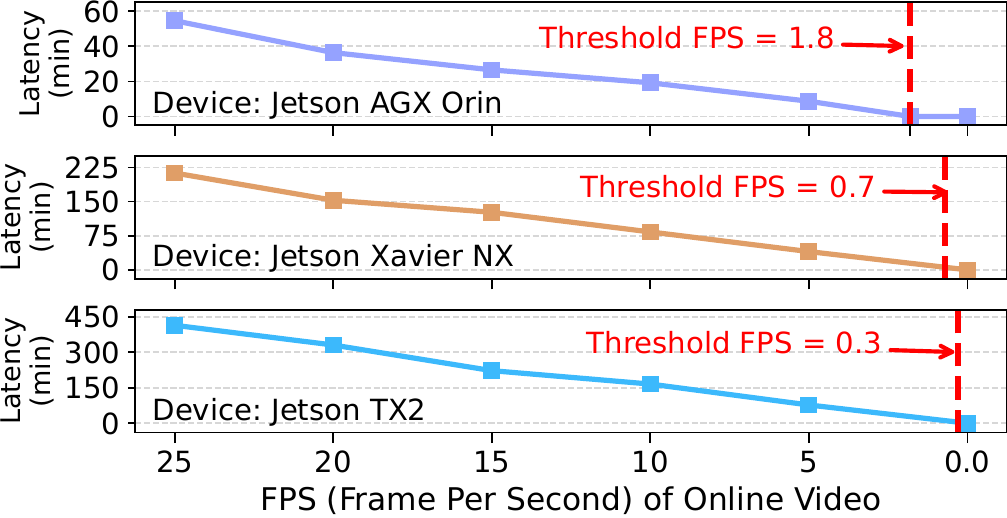}
    \caption{Embedding latency versus FPS across edge devices, with the threshold marking the maximum FPS each device can sustain for real-time embedding.}
    \label{fig:motivation_fps}
    \vspace{-15pt}
\end{figure}

\begin{figure*}[t]
    \centering
    \setlength{\abovecaptionskip}{-0.1cm}
    \subfigure[Accuracy on Video-MME (Short).]{
        \includegraphics[width=0.315\textwidth]{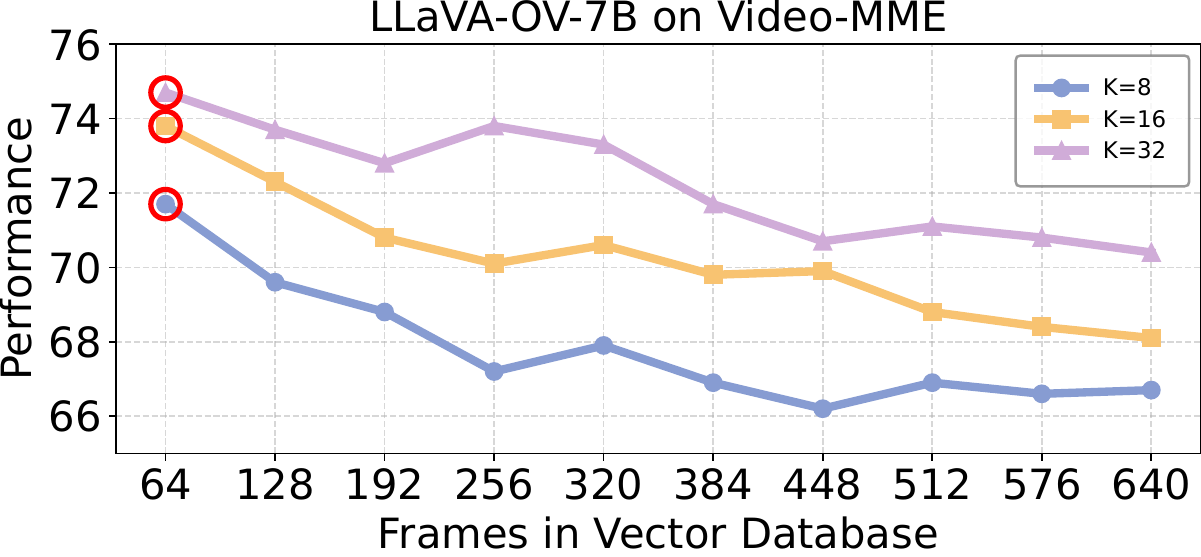}
        \label{fig:rag_frames}
    }
    \hfill
    \subfigure[Frame-wise similarity scores.]{
        \includegraphics[width=0.315\textwidth]{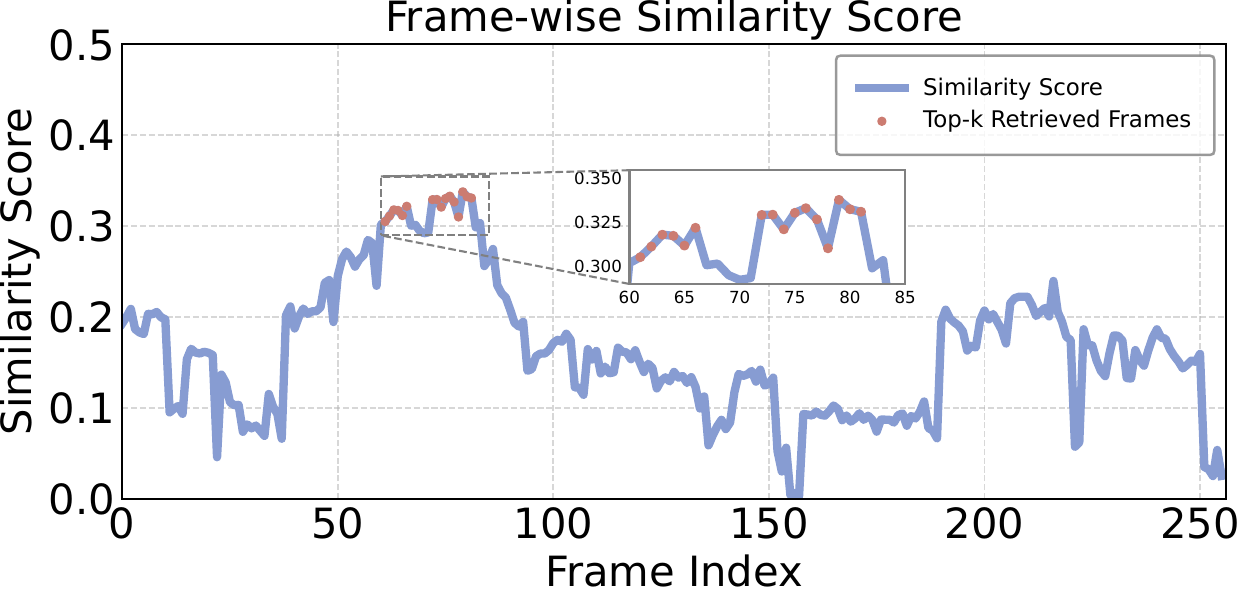} 
        \label{fig:relevance_curve}
    }
    \hfill
    \subfigure[Top-16 frames based on similarity scores.]{
        \includegraphics[width=0.315\textwidth]{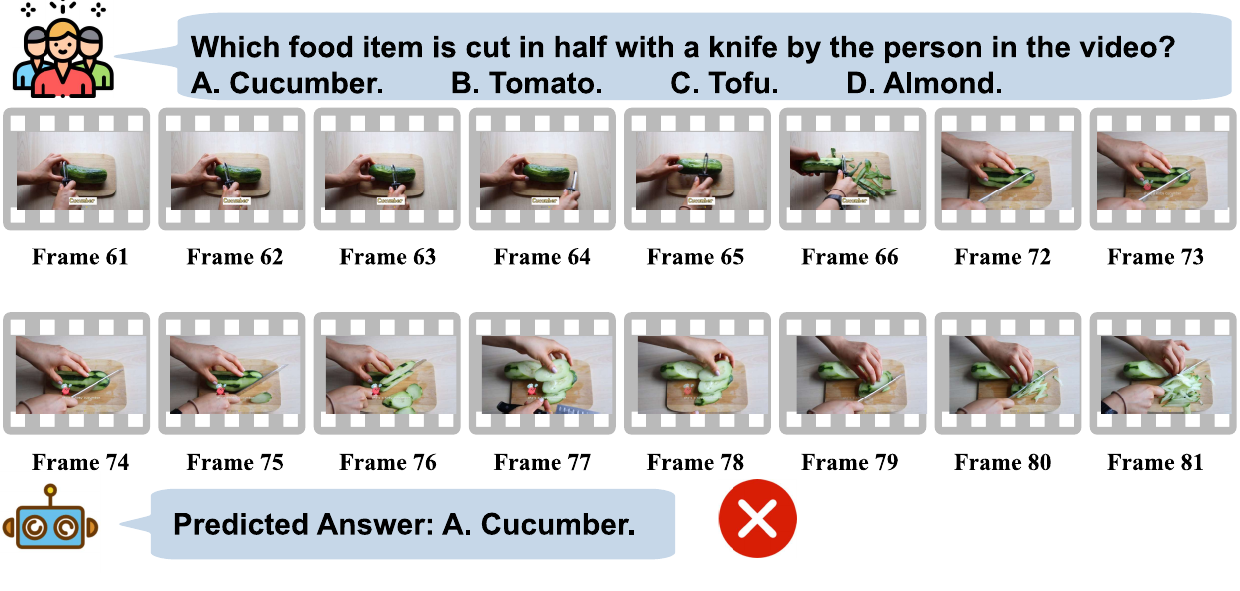}
        \label{fig:Topk}
    }
    \caption{
    \textbf{(a) Accuracy on the short split of Video-MME (w/o subtitles).} 
    Adding redundant frames to the vector database leads to performance degradation, highlighting the importance of diversity in frame selection. 
    \textbf{(b) Visualization of frame selection using Top-K retrieval.} 
    The blue curve shows similarity scores across 256 uniformly sampled frames; red dots indicate selected frames, which are concentrated around adjacent timestamps. 
    \textbf{(c) Retrieved frames using Top-K selection.} 
    All selected frames focus exclusively on the “cucumber” segment, ignoring other answer options and resulting in incorrect predictions.
    } 
    \label{fig:retrieval_overview}
    \vspace{-15pt}
\end{figure*}

\section{Edge-Cloud Disaggregated Architecture for Online Video Understanding}
\label{sec:ec-dis}

\subsection{Opportunities for Online Video Understanding}
\label{sec:opportunity}
To achieve our design goals, we identify two key components that offer strong potential for building our system.
\subsubsection{Multimodal Embedding Models}
Embedding was originally introduced to transform data such as text, images, or audio into vector representations, enabling similarity comparisons within the same modality \cite{lin2017structured, dosovitskiy2020image}. However, unimodal embedding is inherently limited in capturing cross-modal information.
To bridge this gap, multimodal embedding models (MEMs) \cite{zhou2024megapairs, radford2021learning, zhai2023sigmoid} have been developed to map different modalities into a shared representation space.
By optimizing the alignment of representations across modalities\footnote{This work focuses on visual and linguistic modalities, specifically image and text, for video understanding. Although other modalities (e.g., audio, temperature signal) are not included, \texttt{Venus} can be naturally extended by integrating MEMs that support additional modalities, such as ImageBind \cite{girdhar2023imagebind}.}, MEMs facilitate semantic interaction between modalities through a shared embedding space. In the context of online video understanding, this enables user-friendly querying via natural language to efficiently retrieve query-relevant video frames.

\subsubsection{Vector Database}
MEMs enable embedding multimodal data into a unified vector space, where vector databases enable efficient storage and fast nearest neighbor retrieval for similarity search \cite{wang2025timestamp}.
Given a set of data vectors $\{x_i\}$ and a query $q$, the goal is to identify the most similar vectors to $q$, where similarity is typically defined using either Euclidean distance or inner-product.
Facebook AI Similarity Search (FAISS) \cite{johnson2019billion} exemplifies such systems, accelerating nearest neighbor search via clustering, quantization, graph-based indexing, and hardware acceleration for scalable, low-latency retrieval in large vector spaces.

\subsection{An Edge-Cloud Disaggregated Architecture}
In online video understanding, streaming video is typically generated on edge devices, whereas VLMs, due to their high resource demands, are commonly hosted in the cloud.
However, fully uploading relevant video frames to cloud-hosted VLM service can incur a huge volume of bandwidth usage as well as on-cloud computation overhead, as demonstrated in \S \ref{sec:motivation-exp}.
To practically integrate cloud-hosted VLM regarding our design goals (\S \ref{sec:design-goal}), we must alleviate this edge-cloud communication bottleneck and reduce redundant data transmission as much as possible. 
Motivated by the opportunities in \S \ref{sec:opportunity}, we propose an edge–cloud disaggregated architecture, where the edge side applies a group of memory-and-retrieval modules to intelligently select critical video frames, and the cloud is dedicated to VLM reasoning tasks.

Figure \ref{fig:strawman} illustrates our proposed workflow, comprising two main stages: ingestion and querying.
In the ingestion stage, the camera continuously captures video streams at a fixed frame rate. Each video frame is then processed by a MEM to produce a vector representation within a unified image-text embedding space.
Subsequently, these vector representations are inserted into a vector database, where index structures are constructed to enable accelerated similarity search.
In the query stage, a user query in natural language is processed by the same MEM and mapped into the unified multimodal embedding space. The resulting query vector is then compared against the stored indexed vectors in the vector database via similarity computation, and the Top-K most relevant video frames are retrieved. These retrieved frames are subsequently transmitted to a cloud-hosted VLM service for inference, which produces the final reasoning results.

\begin{figure*}[t!]
\setlength{\abovecaptionskip}{-0.1cm}
	\centering
	\includegraphics[width=0.9\textwidth]{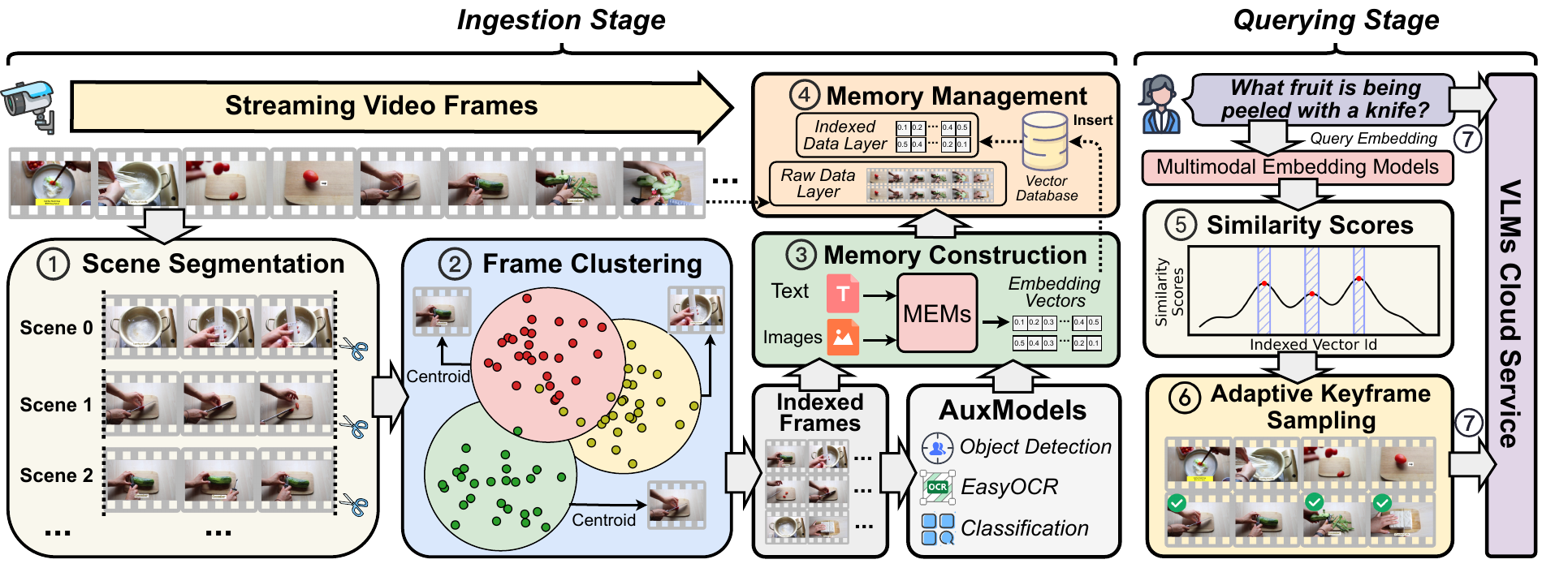} %
	\caption{The system overview of our proposed \texttt{Venus}, which consists of an ingestion stage and a querying stage.}
	\label{fig:system_overview}
    \vspace{-15pt}
\end{figure*}

\subsection{Technical Challenges in Practical Deployment}
The aforementioned architecture provides a straightforward design for online video understanding applications powered by VLMs.
While conceptually simple, it suffers from several key challenges that hinder real-world deployment.

\noindent\textbf{\circledwhite{1} High latency hinders real-time embedding and ingestion.}
Typical edge cameras capture video at 25–60 FPS \cite{TL_IPC632P_A4}, which imposes substantial overhead when applying MEMs for frame-wise embedding, particularly on resource-constrained edge devices. When the input FPS exceeds the device’s embedding throughput, frames accumulate in a backlog. Upon receiving a user query at timestamp $t$, the system must first embed all pending frames before $t$ before performing retrieval, introducing latency and degrading responsiveness.

To assess the feasibility of real-time on-device embedding, we evaluate three representative edge platforms: NVIDIA Jetson AGX Orin \cite{jetson-orin}, Jetson Xavier NX \cite{jetson-xavier}, and Jetson TX2 \cite{jetson-tx2} under realistic streaming conditions, where frames are processed on the fly as they are captured.
As shown in Figure \ref{fig:motivation_fps}, embedding latency increases rapidly with higher FPS, making real-time embedding infeasible beyond a certain FPS threshold. At the native frame rate of typical network cameras (e.g., 25 FPS), embedding delay exceeds 212 minutes, which is clearly impractical for online query. Our results indicate that real-time embedding and ingestion are only achievable when the frame rate is drastically reduced, such as 0.3 FPS for Jetson TX2, 0.7 FPS for Jetson Xavier NX, and 1.8 FPS for Jetson AGX Orin. However, such aggressive uniform downsampling leads to severe information loss. If key frames required to answer user queries are discarded, the system may entirely fail to perform accurate video understanding.

\noindent\textbf{\circledwhite{2} Excessively redundant frames overwhelm the memory database and degrade retrieval performance.}
Embedding and retaining all video frames may seem beneficial for preserving fine-grained details and improving the chances of accurate retrieval. 
However, our analysis shows that excessive redundancy in the memory database degrades both retrieval performance and the accuracy of VLM-based reasoning. To evaluate this effect, we conduct experiments on the Video-MME short split using uniform sampling with adjustable intervals, which controls the number of frames embedded and stored in the database. Counterintuitively, the highest accuracy is often achieved when only 64 frames are retained, as illustrated in Figure \ref{fig:rag_frames}.
We use a case study to further investigate the underlying reason, as shown in Figure \ref{fig:relevance_curve} and Figure \ref{fig:Topk}. The results reveal that the memory database contains numerous temporally adjacent and visually similar frames, leading the retrieval algorithm to return redundant, near-duplicate content. 
Moreover, redundant frames also increase indexing complexity and storage overhead of database.

\noindent\textbf{\circledwhite{3} Lack of diversity and adaptivity in Top-K selection algorithm.}
Although simple and effective for direct visual matching, Top-K greedy selection exhibits notable limitations. 
First, Top-K sampling often selects redundant frames that are temporally adjacent, especially when the database is densely populated with frames from a single scene. As a result, the retrieved frames may lack contextual diversity and fail to capture the scene coverage necessary for accurate video question answering.
As shown in Figure \ref{fig:relevance_curve}, the Top-K selected frames are heavily concentrated around frame index 75, while relevant scenes between indices 190 and 210 are largely overlooked due to the greedy nature of Top-K selection.
Second, Top-K selection requires a manually specified, static $K$ value, which is difficult to adapt to the complexity and variability of real-world online query scenarios. Different queries, network conditions, and latency requirements call for adaptive adjustment of both the number of sampled frames and the scope of scene coverage.

\section{Venus System Design}
\subsection{Venus System Overview}
In this paper, we introduce \texttt{Venus}, an on-device memory-and-retrieval system for efficient online video understanding. \texttt{Venus} builds upon our previously proposed edge–cloud disaggregated architecture to further address the technical challenges of practical deployment.
Figure \ref{fig:system_overview} illustrates its two-stage workflow:
In the \textit{ingestion stage}, streaming video from the edge camera is processed by the scene segmentation (Step \circledwhite{1}) and frame clustering (Step \circledwhite{2}) modules to filter out redundancy.
The centroids of the resulting clusters are selected as indexed frames and embedded into vector representations using MEMs, supplemented by lightweight auxiliary models, to construct a sparse memory index (Step \circledwhite{3}).
Indexed vectors and raw frames are organized into a hierarchical memory to support accurate and efficient recall and retrieval (Step \circledwhite{4}).
In the \textit{querying stage}, the user query is encoded using the same MEM and matched against indexed vectors to compute similarity scores (Step \circledwhite{5}).
An adaptive keyframe sampling algorithm is employed to ensure both the relevance and diversity of the selected frames (Step \circledwhite{6}).
Finally, the user query and the selected keyframes are sent to the cloud-hosted VLM service to perform fast and accurate reasoning (Step \circledwhite{7}).

\subsection{Scene Segmentation and Clustering for Frame Filtering}
\label{sec:scene_clustering}
\begin{figure}[t!]
    \setlength{\abovecaptionskip}{-0.1cm}
	\centering
	\includegraphics[width=0.9\columnwidth]{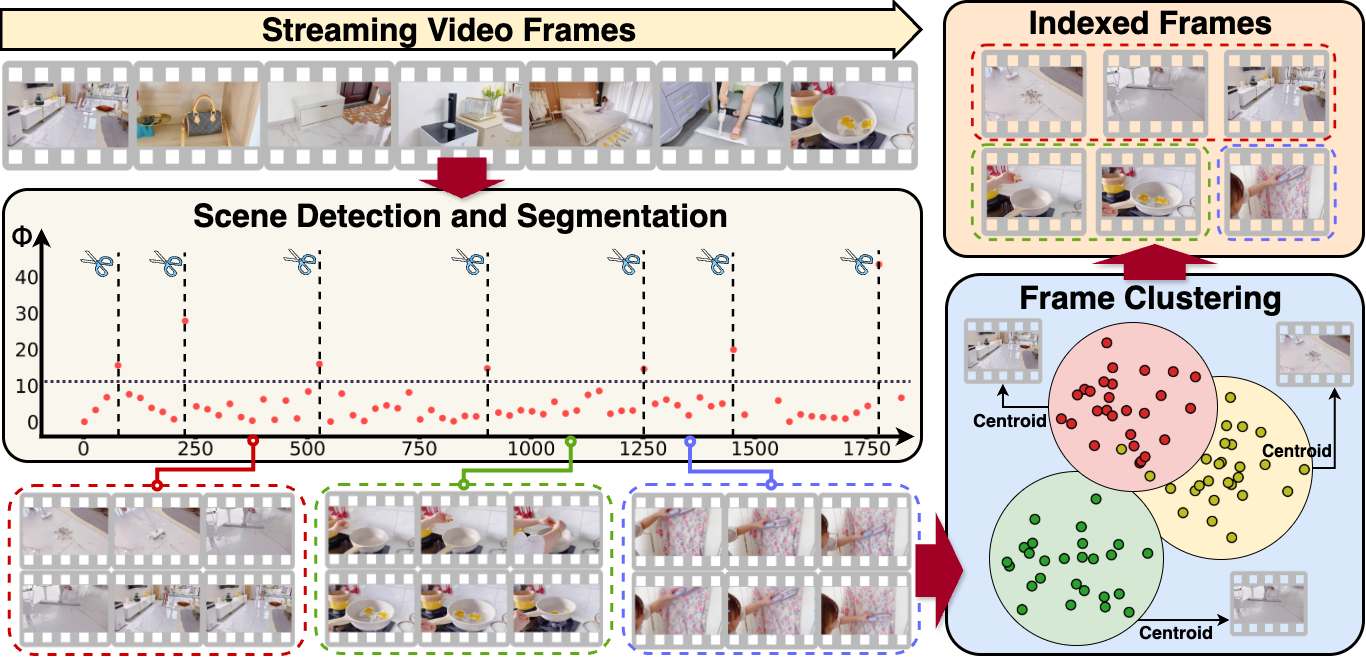} %
	\caption{An illustration of our scene segmentation and frame clustering module.}
	\label{fig:cluster}
    \vspace{-15pt}
\end{figure}

Streaming video at 25–60 FPS \cite{TL_IPC632P_A4} on typical edge cameras makes frame-wise perception and memory prohibitively expensive for real-time processing.
Uniform downsampling is widely used in prior work, but often results in severe and uncontrollable information loss.
\texttt{Venus} aims not to prune frames aggressively, but instead to preserve raw video and construct a sparse index efficiently using lightweight online algorithms.
The index comprises representative frames that summarize visually similar content, enabling efficient localization of query-relevant scenes and subsequent fine-grained sampling during the querying stage.
This section focuses on selecting indexed frames through a scene segmentation and frame clustering module, which consists of two steps: scene segmentation and frame clustering, as illustrated in Figure \ref{fig:cluster}.

\subsubsection{Scene Detection and Segmentation}
\texttt{Venus} takes streaming video as input and incrementally partitions it into contiguous temporal segments using a coarse-grained algorithm, termed scene detection and segmentation.
We perform a lightweight coarse-grained partitioning of the video stream using visual features including hue, saturation, brightness, and edge maps \cite{cao2006density}.
Specifically, we define a frame-level metric $\phi$ to quantify visual differences between consecutive frames based on intrinsic pixel-level features. Given a streaming video $\mathcal{F} = \{f_0, f_1, \dots\}$, where $f_i$ denotes the frame at time step $i$, the scene tracking score for $f_i$ is computed as: 
\begin{equation}
\begin{aligned}
\phi(f_i) &= \frac{
    \left\| \mathbf{w} \odot (\mathbf{v}_i - \mathbf{v}_{i-1}) \right\|_1
}{
    \left\| \mathbf{w} \right\|_1
}, \\
\text{where} \quad
\mathbf{v}_i &= [H(f_i),\, S(f_i),\, L(f_i),\, E(f_i)], \\
\mathbf{w} &= [w_H,\, w_S,\, w_L,\, w_E].
\end{aligned}
\end{equation}
Here, $H(\cdot)$, $S(\cdot)$, $L(\cdot)$, and $E(\cdot)$ denote the hue, saturation, lightness, and edge feature maps of each frame, respectively. 
$||\cdot||_1$ is the L1 norm.
The weights $\{w_H, w_S, w_L, w_E\}$ control the contribution of each modality to the overall score.
A scene boundary is detected when $\phi(F_i)$ exceeds a predefined threshold, segmenting the video into discrete partitions.
For video streams with minimal scene transition, such as fixed-view cameras, coarse-grained segmentation may be ineffective. 
To address this, we apply a minimum temporal threshold: if no scene change occurs within a set duration, all frames in that period are grouped into a single partition.
Scene partitions $\mathcal{M} = \{m_0, m_1, \dots\}$ are continuously generated and enqueued. Frame clustering module then dequeues each partition for further processing.

\subsubsection{Frame Clustering}
We perform frame clustering within each scene partition for two main reasons: First, grouping visually similar frames at a finer granularity further reduces redundancy. Second, cluster centroids serve as representative frames that summarize similar content, making them natural candidates for building a sparse index.
Existing research \cite{tang2025adaptive} on video analysis also adopts clustering algorithms to reduce redundancy, using methods such as K-Means, DBSCAN, and their variants to group similar frames.
However, a limitation of such clustering algorithms is that frames within a cluster may be temporally disjoint, potentially hindering temporal reasoning.
To this end, \texttt{Venus} adopts a straightforward incremental clustering strategy, which has been well-studied in the literature \cite{cao2006density, o2002streaming}.
Specifically, for each scene partition $m_i \in \mathcal{M}$, clustering proceeds as follows: the first frame is assigned to the initial cluster $c_1$. To cluster a new frame $f_i$, we first flatten its raw pixel values into a vector representation. We then compute the $L_2$ distance between this vector and the centroid of each existing cluster. If the minimum distance is within a threshold, $f_i $ is assigned to the nearest cluster; otherwise, a new cluster is created with $f_i $ as its centroid.
After clustering, we obtain a set of clusters $\mathcal{C} = \{c_0, c_1, ...\}$, from which the centroid frame of each cluster is selected as an indexed frame. These selected indexed frames, being highly sparse compared to the original video stream, are used for embedding and memory index construction, enabling real-time on-device perception.

\subsection{Memory Construction and Management}
\subsubsection{Memory Construction with MEMs}
During streaming video processing, we continuously obtain the cluster centroids through scene segmentation and clustering, resulting in a set of indexed frames $\mathcal{K} = \{ k_0, k_1, \dots, k_n\}$. These indexed frames are further used to construct the index for the video memory.
Current MEMs, while effective at extracting visual features, remain limited in their ability to recognize characters, perform object recognition, and identify actions, often underperforming compared to lightweight proprietary models.
To better leverage visual information and mitigate hallucinations during user query retrieval, we incorporate lightweight auxiliary models to preprocess video frames and construct auxiliary prompts that enhance indexing accuracy:
\begin{equation}
    \mathcal{T} = \{ t_i = \mathrm{AuxModels}(k_i) \mid k_i \in \mathcal{K} \}
\end{equation}
$\mathrm{AuxModels}$ is a collection of lightweight proprietary models and algorithms, such as OCR \cite{easyocr} and YOLO \cite{redmon2016you}, which can be dynamically configured according to application needs and edge device constraints. Their outputs are formatted into predefined textual templates to construct auxiliary prompts. 
Finally, the indexed frames and their associated auxiliary prompts are fed into the MEMs to generate vector representations within a unified image-text embedding space:
\begin{equation}
    \mathcal{O} = \{ o_i = \mathrm{MEM}(k_i, t_i) \mid k_i \in \mathcal{K}, t_i \in \mathcal{T} \}.
    \label{equ:vector}
\end{equation}

\begin{figure}[t!]
    \setlength{\abovecaptionskip}{-0.1cm}
	\centering
	\includegraphics[width=0.9\columnwidth]{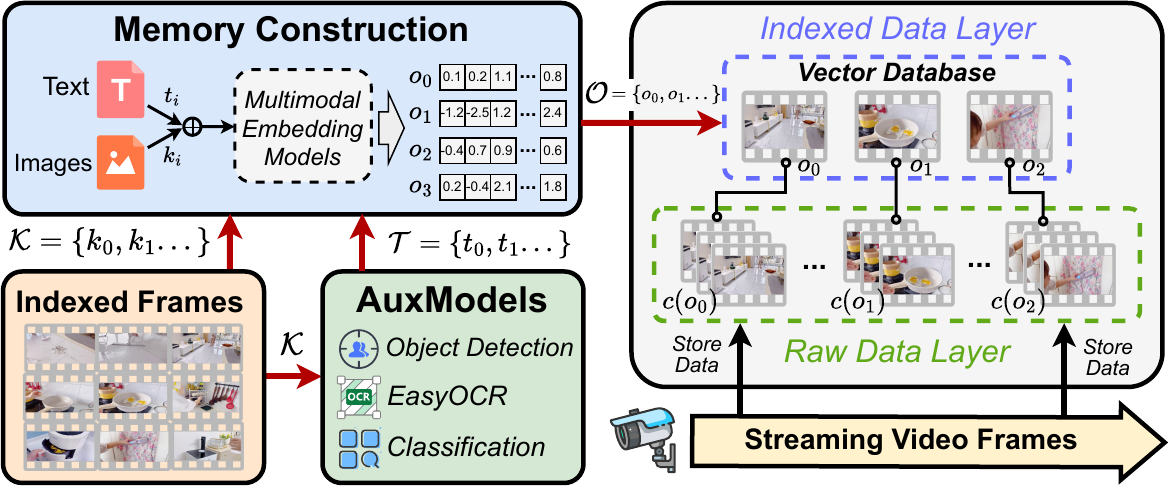} %
	\caption{An illustration of our memory construction and management module.}
	\label{fig:memory}
    \vspace{-15pt}
\end{figure}

\subsubsection{Hierarchical Memory Management with Semantic Indexing}
Our memory management architecture is organized into two hierarchical layers: the \textit{raw data layer} and the \textit{index data layer}. 
Figure \ref{fig:memory} illustrates the workflow of our memory construction and management module.
The \textit{raw data layer} stores streaming video frames captured by edge cameras in their original, unprocessed format. It serves as a persistent archive of historical data and provides a reliable source for accurate user query reasoning.
The \textit{index data layer} builds a structured and semantic index over the stored content, enabling query-driven retrieval and adaptive memory access with high semantic fidelity. Specifically, we insert the vector representations of indexed frames, derived from Equation \ref{equ:vector}, into a high-performance local vector database. Each indexed vector $o_i$ is linked to its corresponding scene cluster $c(o_i)$ in the raw data layer, enabling efficient similarity-based retrieval and adaptive sampling based on query semantics.
\textit{This hierarchical memory management is cognitively inspired by how the human brain recalls memories: first efficiently locating relevant scenes, and then reconstructing finer details.}

\subsection{Query-Relevant Keyframe Retrieval}
\label{sec:Online_Keyframes_Selection}
\subsubsection{Sampling-Based Diversity-Preserving Frame Retrieval}
The design of the ingestion stage supports real-time on-device perception of video streams and hierarchical memory construction.
Upon receiving a user query, \textit{Venus} enters the querying stage, retrieving semantically relevant frames from on-device memory and forwarding them to the cloud-hosted VLMs for reasoning.

\textit{Venus} first embeds the user query $Q$ using the same MEM model and computes its similarity scores with sparsely indexed frames in the vector database, typically using cosine similarity:
\begin{equation}
\mathcal{S} = \left\{ s_i = \cos\left( \mathrm{MEM}(Q),\ o_i \right) \,\middle|\, o_i \in \mathcal{O} \right\}.
\end{equation}
Rather than greedily selecting the Top‑K frames with the highest similarity scores from $\mathcal{S}$, we adopt a sampling-based strategy to ensure both relevance and diversity. 
Specifically, we construct a discrete, query-guided probability distribution over indexed vectors in a vector database based on their similarity scores to the query:
\begin{equation}
\mathcal{P} = \left\{ p_i = \frac{e^{s_i / \tau}}{\sum_{j=1}^{k} e^{s_j / \tau}} \,\middle|\, s_i \in \mathcal{S} \right\},
\end{equation}
where $\tau$ is a temperature parameter controlling the sharpness of the distribution.
Next, we sample $N$ times from the discrete distribution $\mathcal{P}$, resulting in a count function $n : \mathcal{O} \rightarrow \mathbb{N}$, where $n(o_i)$ denotes the number of times indexed vector $o_i$ is selected. For each selected indexed vector, we uniformly sample $n(o_i)$ frames from its associated scene cluster $c(o_i)$, promoting diverse coverage within a cluster.
The sampling-based frame retrieval assigns higher selection probabilities to query-relevant frames while preserving non-zero probabilities for others, striking a balance between relevance and contextual-temporal diversity.
Additionally, by controlling the distribution sharpness through the parameter $\tau$, we provide flexible adjustment of the relevance-diversity trade-off.

\begin{figure}[t!]
    \setlength{\abovecaptionskip}{-0.1cm}
    \centering
    \includegraphics[width=0.9\linewidth]{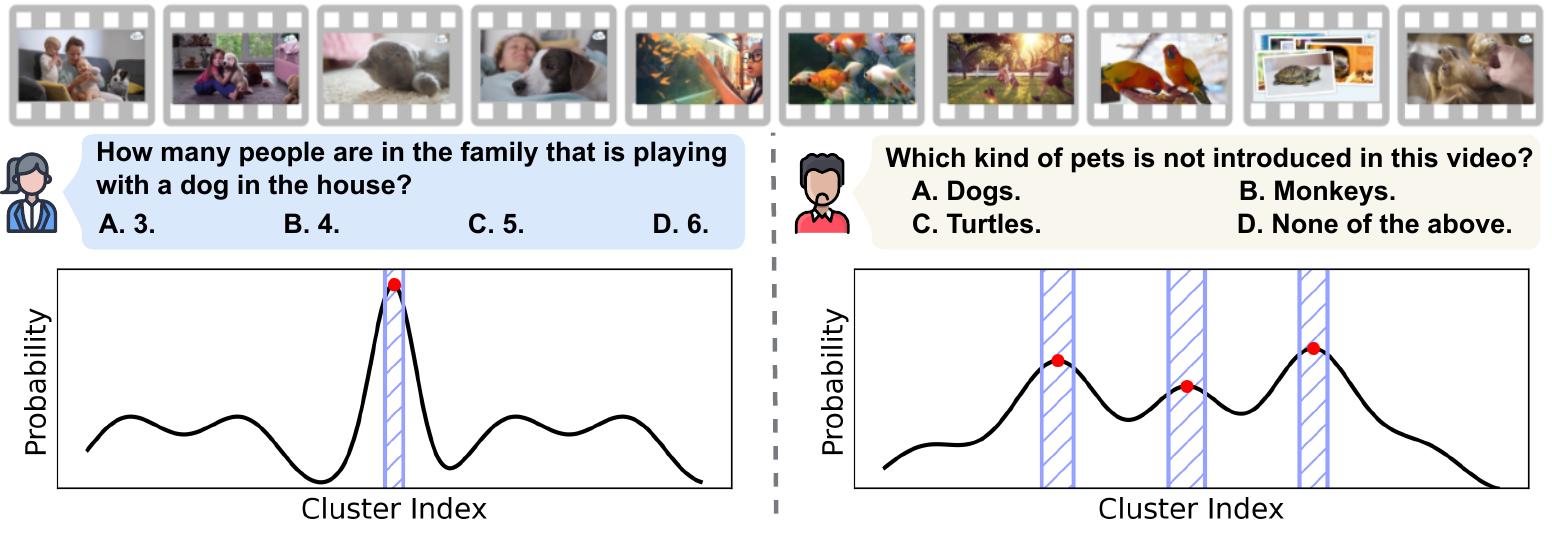}
    \caption{Different types of queries yield different probability distributions.}
    \label{fig:sample}
    \vspace{-15pt}
\end{figure}

\subsubsection{Adaptive Keyframe Retrieval with Progressive Sampling}
Sampling-based retrieval requires a predefined selection budget $N$. However, setting a fixed budget is challenging: a uniformly large selection budget $N$ can lead to redundant frame retrieval, increasing transmission latency and VLMs inference cost, while a small $N$ may fail to meet the accuracy needs of diverse query types.
We conduct a case study to analyze how different types of queries yield distinct probability distributions $\mathcal{P}$.
As shown in the Figure \ref{fig:sample}, when relevant frames are concentrated in a narrow temporal region, only a few samples are sufficient to cover the necessary visual evidence for reasoning.
In contrast, when relevant frames are dispersed across multiple regions, more samples are required to ensure sufficient diversity and coverage.

Motivated by this observation, we design an adaptive keyframe retrieval (\textit{AKR}) using a threshold-driven progressive sampling algorithm.
Specifically, \textit{AKR} progressively samples indexed vectors from the discrete distribution $\mathcal{P}$, maintaining a set $\mathcal{I}$ of selected indices. The sampling process terminates once the cumulative probability of selected frames satisfies:
\begin{equation}
\frac{\sum_{j \in \mathcal{I}} p_j}{\beta} \geq \theta,
\end{equation}
where $\theta$ is a predefined threshold (e.g., $90\%$), thereby avoiding redundant frame transmission and excessive VLMs inference.
To prevent premature termination, we introduce a lower bound on the number of frames to sample, controlled by a parameter $\beta$. The minimum number of retrieved frames is computed as: 
\begin{equation}
N_{\text{min}} = \beta \cdot \left\lceil \theta/\max\limits_{p_j \in \mathcal{P}} p_j\right\rceil.
\end{equation}
The upper bound on the number of sampled frames $N_{\text{max}}$ is constrained by the maximum tolerable transmission delay, which is determined by the bandwidth of the edge network.

\section{IMPLEMENTATION AND EVALUATION}
% We have fully implemented the prototype system of \texttt{Venus} and baselines with $\sim$1000 LoC in Python atop PyTorch. 
% This section presents a comprehensive evaluation of the \texttt{Venus} prototype, including comparisons with baselines across various datasets, models, and edge platforms.

\subsection{Experimental Setup}

\subsubsection{Hardware Setup} 
On the edge side, we used NVIDIA Jetson AGX Orin \cite{jetson-orin} devices for our experiments with each device equipped with an NVMe SSD for external storage.
On the cloud side, we use a server with NVIDIA L40S GPUs (48 GB memory) to emulate cloud compute resources, where the VLMs are deployed. The network bandwidth between the edge and cloud is fixed at 100 Mbps, which is typical for wired or wireless edge networks.

\newcolumntype{C}[1]{>{\centering\arraybackslash}p{#1}}

\begin{table*}[ht]
\setlength{\abovecaptionskip}{-0.1cm}
 \setlength{\belowcaptionskip}{-0.4cm}
\setlength{\tabcolsep}{1pt}
\centering
\caption{Comparison with query-irrelevant baselines across different datasets, VLMs, and frame sampling budgets.}
\label{tab:query_irrelevant}
% \begin{tabular}{lcccccccccccccc}
\begin{tabular}{C{2.0 cm} C{2.2cm} 
                C{1.0cm} C{1.0cm} C{1.0cm} C{1.0cm} 
                C{1.0cm} C{1.0cm} C{1.0cm} C{1.0cm} 
                     C{1.2cm} C{1.2cm} C{1.2cm} C{1.2cm} }
                % C{1.1cm} C{1.1cm} C{1.1cm} C{1.1cm}}

\Xhline{1.2pt}
\addlinespace[2pt]
\multirowcell{2}[-3pt][c]{\textbf{Model}} & 
\multirowcell{2}[-3pt][c]{\textbf{Method}} & 
\multicolumn{4}{c}{\textbf{Video-MME (N=16)}} &
\multicolumn{4}{c}{\textbf{Video-MME (N=32)}} &
\multicolumn{2}{c}{\textbf{EgoSchema (N=16)}} &
\multicolumn{2}{c}{\textbf{EgoSchema (N=32)}}
\\
\cmidrule(lr){3-6} 
\cmidrule(lr){7-10} 
\cmidrule(lr){11-12} 
\cmidrule(lr){13-14} 
\multicolumn{2}{c}{} & Short & Medium & Long & Overall 
  & Short & Medium & Long & Overall 
  & Full & Subset & Full & Subset \\

\midrule
\midrule
 \multirow{4}{*}{LLaVA-OV-7B} & Uniform Sampling  
 &61.2  &50.4   & 45.9 & 52.5  
 & 63.1  & 52.3 & 48.6 & 54.6
&50.2 & 51.8
&50.5  & 51.2\\
 
& MDF 
&59.1&47.9  & 44.6 & 50.5
&61.0 & 48.8 & 44.6 & 51.5
&49.8&50.4
 & 51.6 & 52.2 \\

& Video-RAG 
&61.5 & 49.6 & 43.7 &  51.6
& 63.7  & 51.3 & 46.3& 53.8
&49.3& 52.8
& 49.7 & 51.4 \\
  
  \rowcolor{rowcolor}
  \cellcolor{white} & \texttt{Venus}  
  &\textbf{64.4} &\textbf{ 56.2}  & \textbf{51.0} &\textbf{57.2}
  & \textbf{67.4}  &\textbf{ 59.6}&\textbf{56.4}&\textbf{ 61.1}
  &\textbf{52.8}& \textbf{55.0 }
  &\textbf{53.3}   & \textbf{55.5}  \\
 
\Xcline{2-14}{0.8pt}
\addlinespace[2pt]

 \multirow{4}{*}{ Qwen2-VL-7B}  & Uniform Sampling 
 &  64.7  &  50.6   & 45.9 &  53.7
 & 68.0     & 54.1  &48.6 & 56.9
 & 64.4 & 60.6
  &63.8 & 61.4\\
 
  & MDF 
  &  64.1  & 50.2  & 46.2  & 53.5 
  & 67.8   &53.1 & 47.3& 56.1
 &59.2 & 58.8 
  & 61.4 &62.2 \\
  
  &Video-RAG 
  &67.1 & 50.8  & 46.3 & 54.7
  & 69.2   & 55.1 & 48.7& 57.7
  & 63.7&60.3
  & 63.6 & 60.6  \\
   \rowcolor{rowcolor}
   
\cellcolor{white} & \texttt{Venus}
& \textbf{70.6} & \textbf{56.8} &\textbf{50.2} &\textbf{59.2}
& \textbf{74.3}  & \textbf{63.8} &\textbf{53.6}& \textbf{63.9}
& \textbf{68.9} & \textbf{63.8}
&\textbf{69.5}&  \textbf{65.7}   \\

\Xhline{1.2pt}

\end{tabular}

\vspace{-10pt}
\end{table*}

\begin{table*}[ht]
\setlength{\tabcolsep}{8.2pt}
\setlength{\abovecaptionskip}{-0.1cm}
 \setlength{\belowcaptionskip}{-0.4cm}
\centering
\caption{Comparison with query-relevant baselines across different datasets, VLMs and deployment strategies.}
\label{tab:query_relevant}
\begin{tabular}{lcc cc cc cc cc}
\Xhline{1.2pt}
\addlinespace[2pt]
\multirowcell{2}[-3pt][c]{\textbf{Model}} & 
\multirowcell{2}[-3pt][c]{\textbf{Method}} & 
\multicolumn{2}{c}{\textbf{Video-MME (Short)}} & 
\multicolumn{2}{c}{\textbf{Video-MME (Medium)}} & 
\multicolumn{2}{c}{\textbf{Video-MME (Long)}} &
\multicolumn{2}{c}{\textbf{EgoSchema (Full)}}  \\
\cmidrule(lr){3-4} \cmidrule(lr){5-6} \cmidrule(lr){7-8} \cmidrule(lr){9-10}
& & Accuracy & Latency 
  & Accuracy & Latency 
  & Accuracy & Latency 
  & Accuracy & Latency \\

\midrule
\midrule
 \multirow{6}{*}{LLaVA-OV-7B}   & AKS (\textit{Cloud-Only})    
 & 65.9&46.8s   
 & 52.1&2.7min 
 &52.2&11.2min  
 & 51.6 & 78.3s \\

& AKS (\textit{Edge-Cloud})    
 & 65.9&419.1s  
 & 52.1&43.7min  
 &52.2&212.1min 
 & 51.6 &924.0s \\
 
&BOLT (\textit{Cloud-Only})      
&\textbf{67.9} &43.9s   
&55.9&2.5min  
&56.2&10.7min  
& 52.4&70.1s   \\

&BOLT (\textit{Edge-Cloud})     
&\textbf{67.9} &398.1s   
&55.9&41.8min  
&56.2&206.7min
& 52.4&896.9s   \\

& Vanilla     
 &63.6&379.0s   
 &52.5&39.6min  
 &51.0&192.2min  
 &50.8&852.7s \\
 
\rowcolor{rowcolor} \cellcolor{white}  & \texttt{Venus}   
&67.4&\textbf{4.7s}   
&\textbf{59.6}&\textbf{4.9s}  
&\textbf{56.4}&\textbf{5.1s}  
&\textbf{53.3}&\textbf{4.8s}      \\

\Xcline{2-10}{0.8pt}
\addlinespace[2pt]

 \multirow{6}{*}{Qwen2-VL-7B}   & AKS (\textit{Cloud-Only})    
 &72.4&46.8s   
 &62.0 &2.8min  
 &52.1 &11.4min 
 &66.5&82.1s \\

& AKS (\textit{Edge-Cloud})    
 &72.4&417.1s   
 &62.0 &44.8min 
 &52.1 &214.8min 
 &66.5&950.2s \\
 
&BOLT (Cloud-Only)      
&\textbf{75.1}&44.8s   
&63.3&2.7min
&52.8&11.2min  
&67.4&75.9s   
\\
&BOLT (Edge-Cloud)     
&\textbf{75.1}&418.9s   
&63.3&43.0min  
&52.8&212.8min 
&67.4&959.1s  
\\
& Vanilla     
&71.9&391.0s   
&58.7&41.5min 
&50.9&190.9min
&64.0&894.3s  \\
 
\rowcolor{rowcolor} \cellcolor{white}  & \texttt{Venus}   
&74.3&\textbf{4.8s}
&\textbf{63.8}&\textbf{5.1s}  
&\textbf{53.6}&\textbf{5.4s}  
&\textbf{69.5}&\textbf{5.0s}      \\

\Xhline{1.2pt}

\end{tabular}
\vspace{-10pt}
\end{table*}

\subsubsection{Datasets and Models}
We evaluate \texttt{Venus} on two widely used VQA benchmarks, \textit{Video-MME} \cite{fu2025video} and \textit{EgoSchema} \cite{mangalam2023egoschema}, which simulate realistic online video understanding scenarios.
Video-MME consists of 900 videos and 2,700 questions, with some video clips exceeding one hour in length. EgoSchema contains over 5,000 multiple-choice QA pairs grounded in approximately 250 hours of egocentric video footage.
All input videos are uniformly sampled at 8 FPS in our experiments.
We assume that the query arrives immediately after the playback of each video clip, simulating an online video understanding scenario in which users query specific temporal segments of the video in real time.
To evaluate the performance of VLM-based video understanding, we deploy two popular open-source VLMs on the cloud server: LLaVA-OV-7B \cite{li2024llava} and Qwen2-VL-7B \cite{wang2024qwen2}.
We employ 7B models due to hardware constraints, while \textit{Venus}’s modules are independent of the VLM and invoke it only via API calls for final reasoning, enabling access to cloud-hosted models of any size.
We use BGE-VL-large \cite{zhou2024megapairs} as the MEM.
We use EasyOCR~\cite{easyocr} and YOLO~\cite{redmon2016you} as auxiliary models.

\subsubsection{Baselines Methods} 
We compare \texttt{Venus} with existing state-of-the-art VLM-based video understanding methods in terms of accuracy and response latency. The baselines include both approaches without query-relevant frame selection and those incorporating frame selection strategies:
\begin{enumerate}[leftmargin=*, label=(\arabic*)] 
    \item \textit{Uniform Sampling} samples frames at fixed intervals.
    \item \textit{MDF} \cite{han2024self} proposes a query-irrelevant method that adopts a self-adaptive strategy to filter the most dominant frames.
    \item \textit{Video-RAG} \cite{luo2024video} adopts uniform sampling over video frames and incorporates additional auxiliary data to construct a RAG database that enhances VLM performance. 
    \item \textit{AKS} \cite{tang2025adaptive} proposes an adaptive keyframe selection method and employs an optimization algorithm to ensure comprehensive coverage of the selected keyframes.
    \item \textit{BOLT} \cite{liu2025bolt} designs an inverse transform sampling method to prioritize query-relevant frames, improving performance on multi-source retrieval VQA tasks.
    \item \textit{Vanilla} refers to the naive edge–cloud disaggregated architecture introduced in \S \ref{sec:ec-dis}, without any optimization.
\end{enumerate}
For baselines with keyframe selection algorithms, such as AKS and BOLT, we provide two deployment strategies: (1) Both the selection algorithm and VLM inference are executed in the cloud (\textit{Cloud-Only}). (2) Keyframe selection is performed on the edge, and only the selected frames are uploaded to the cloud for inference (\textit{Edge-Cloud}).
\textit{To ensure fair comparison, all \textit{Edge–Cloud} baselines are implemented under an online setting, with preprocessing in the keyframe selection algorithm performed on-the-fly as the video stream progresses.} 

\subsection{Comparison with Query-Irrelevant Baselines}
We compare the accuracy of \texttt{Venus} with baselines without query-relevant frame selection.
We conduct evaluations across different datasets, VLMs, and frame sampling budgets (N=16/32), with results presented in Table \ref{tab:query_irrelevant}. 
The results show that although these methods incur low computational overhead, their query-agnostic nature limits their accuracy on complex video reasoning tasks. In contrast, \texttt{Venus} consistently achieves the highest accuracy across all settings. 
Uniform sampling is particularly problematic for long videos, as it may drop critical frames uncontrollably, leaving VLMs without sufficient visual evidence to answer the query accurately.

% \vspace{-5pt}
\subsection{Comparison with Query-Relevant Baselines}
We evaluate \texttt{Venus} against query-relevant baselines across various datasets, VLMs, and deployment strategies. For fairness, we fix the sampling budget to 32 frames and disable \texttt{Venus}'s adaptive keyframe retrieval. Table \ref{tab:query_relevant} shows that \texttt{Venus} matches or exceeds baseline accuracy while significantly reducing latency. Specifically, our sampling-based retrieval ensures relevance and diversity, achieving parity with AKS and BOLT while outperforming vanilla edge–cloud disaggregated architectures. Regarding efficiency, \texttt{Venus}'s scene segmentation and clustering minimize indexed frames, enabling real-time edge perception. In cloud-only setups, \texttt{Venus} achieves up to $126\times$ speedup on Video-MME Long by avoiding the massive transmission overheads inherent in BOLT and AKS. In edge–cloud deployments, \texttt{Venus} avoids the high on-device computation costs of frame-wise processing, which can result in latencies up to 418.9 seconds on Video-MME Short and scales further with video length.

\subsection{Comparison of Top‑K and Sampling‑Based Retrieval}
\begin{figure}[t!]
    \setlength{\abovecaptionskip}{-0.1cm}
    \centering
    \includegraphics[width=0.9\linewidth]{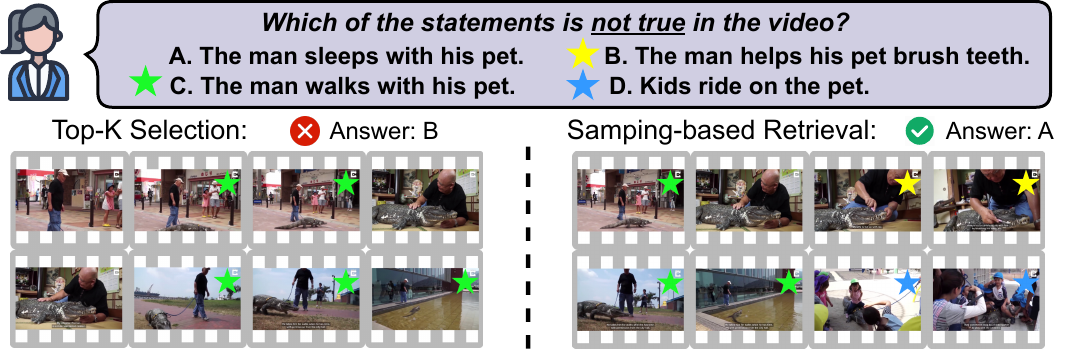}
    \caption{Comparison of vanilla's greedy Top‑K and sampling‑based retrieval.}
    \label{fig:eval-diversity}
    \vspace{-5pt}
\end{figure}

We conducted a case study with a fixed budget of 8 frames to compare the vanilla greedy Top‑K selection with the sampling‑based retrieval adopted by Venus, in order to examine whether sampling can improve the diversity and coverage of the selected results, as illustrated in Figure \ref{fig:eval-diversity}.
We observe that for the same question, the Top‑K selection only covers option C, whereas the sampling‑based retrieval includes content related to options B, C, and D. This broader coverage enables the model to eliminate incorrect options and correctly identify option A as unrelated to the video.

\begin{figure}[t]
    \setlength{\abovecaptionskip}{-0.1cm}
    \centering
    \includegraphics[width=0.95\linewidth]{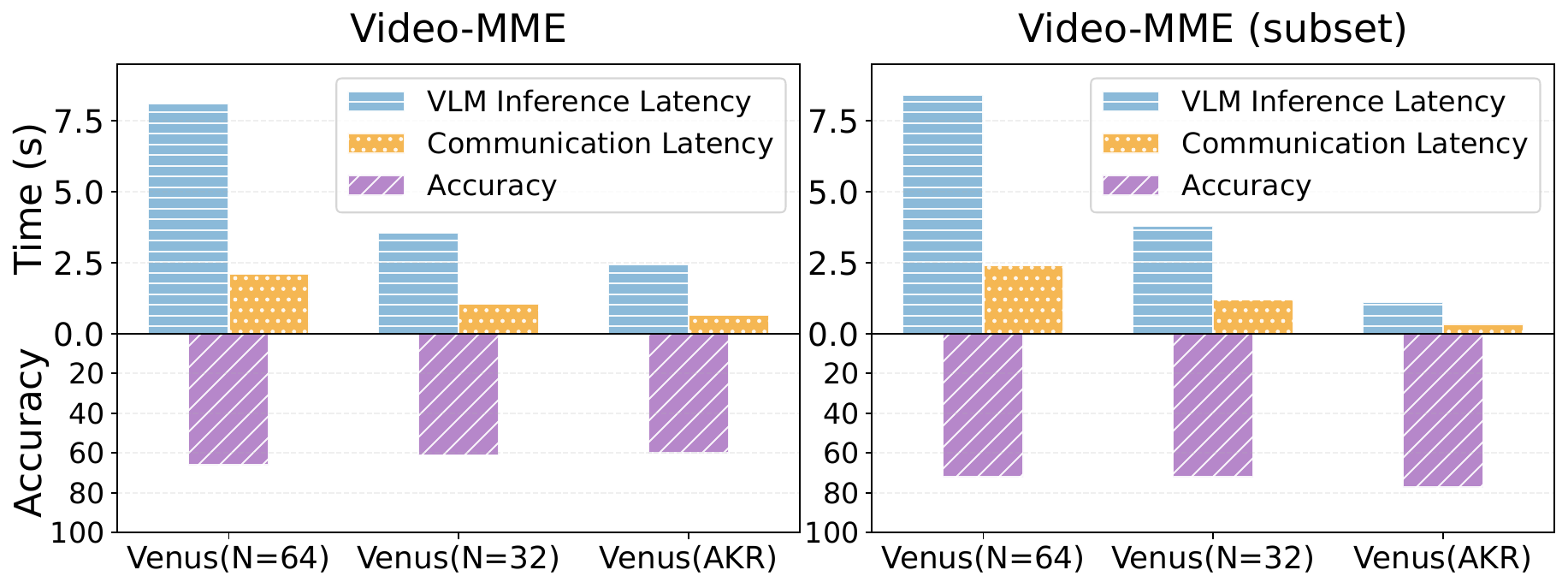}
    \caption{The ablation study on adaptive keyframe retrieval module.}
    \label{fig:ablatio_akr}
    \vspace{-15pt}
\end{figure}

\subsection{Ablation Study on Adaptive Keyframe Retrieval}
% In this section, we evaluate the impact of our Adaptive Keyframe Retrieval (AKR) module.
We compare \texttt{Venus} with AKR ($N_{\text{max}}$=32) against \texttt{Venus} using sampling-based frame retrieval with fixed sampling budgets of 64 and 32 frames, evaluating reasoning accuracy, as well as average inference latency and communication overhead.
We conduct the evaluation on the Video-MME Short dataset. To further highlight the advantage of AKR, we used ChatGPT-4o to help identify 60 queries (\textit{Video-MME subset}) that focus on specific scenes within the video and therefore require only a small number of frames for accurate reasoning, similar to the example shown in Figure \ref{fig:sample}(left). The evaluation results are shown in Figure \ref{fig:ablatio_akr}.
We observe that \texttt{Venus} with AKR achieves accuracy comparable to fixed-budget baselines with 64 or 32 frames on Video-MME, while selecting only about 17 frames on average. This leads to a $1.6\times$–$3.3\times$ reduction in VLM inference and communication overhead.
On the curated subset, AKR further amplifies its advantage, achieving a $3.8\times$–$7.6\times$ reduction in latency. By eliminating redundant frames that may interfere with VLM inference, AKR also outperforms the fixed-budget baselines in accuracy.

\subsection{End-to-end Query Latency Breakdown}
\begin{figure}[t]
    \setlength{\abovecaptionskip}{-0.1cm}
    \centering
    \includegraphics[width=0.9\linewidth]{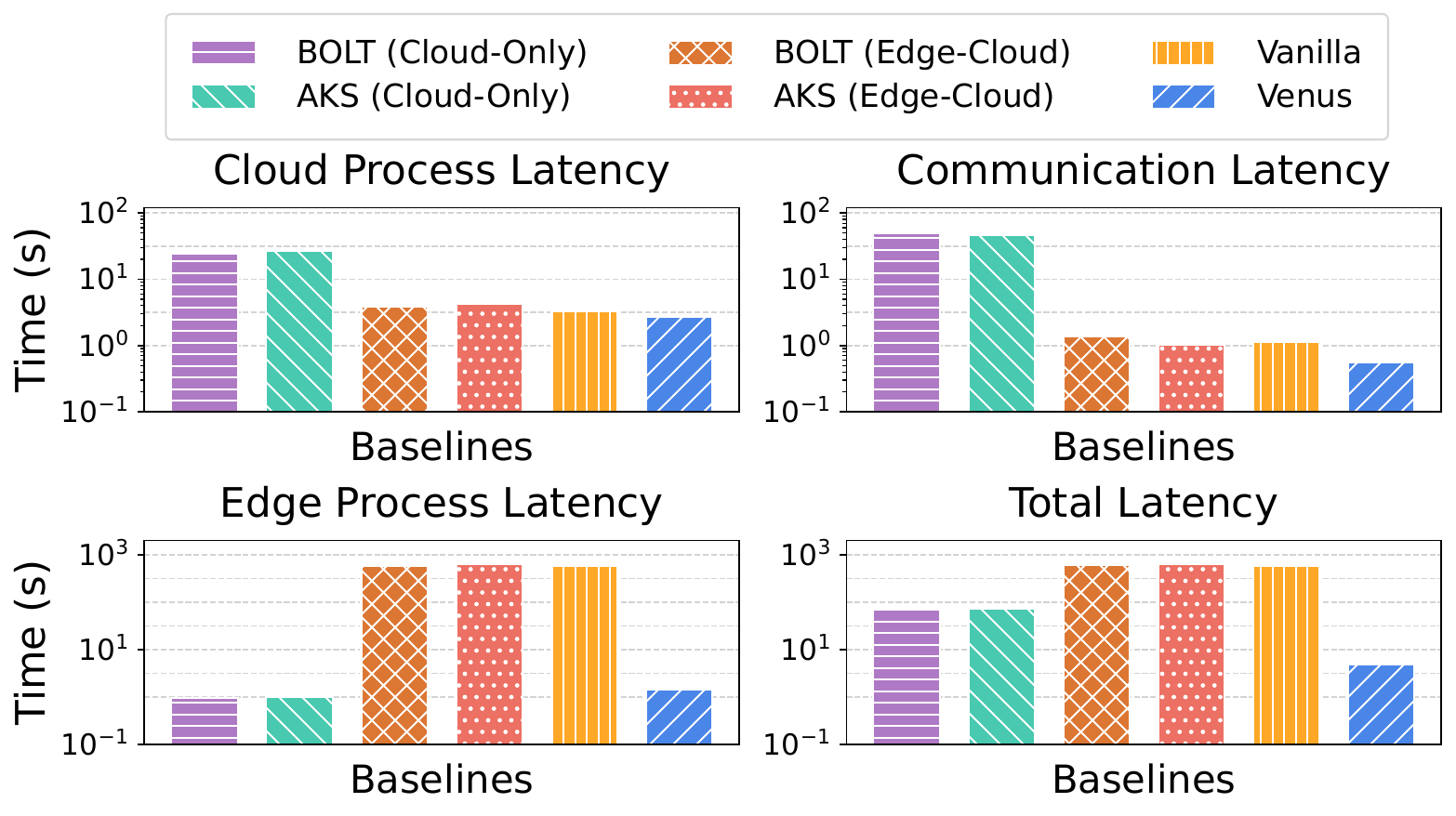}
    \caption{End-to-end query latency breakdown.}
    \label{fig:eval_breakdown}
    \vspace{-15pt}
\end{figure}

We present a breakdown of the end-to-end response latency for both \texttt{Venus} and the baselines, using the Video-MME Short dataset. The average latency of each processing step is reported in Figure \ref{fig:eval_breakdown}.
\texttt{Venus} consistently achieves the lowest latency across all steps, resulting in a $15\times$–$131\times$ speedup in total response latency compared to the baselines.
Specifically, \texttt{Venus} performs scene segmentation and clustering to enable real-time on-device embedding and memory construction, allowing only the query text to be embedded upon query arrival. In addition, \texttt{Venus} leverages AKR to adaptively select keyframes for upload, significantly reducing both communication and cloud-side VLM inference latency.

\section{RELATED WORK}

\noindent \textbf{Pre-VLM Era Online Video Understanding.} 
% Prior to the VLM era, CNNs \cite{he2016deep, terven2023comprehensive} served as the backbone of online video understanding.
% Several studies introduce system-level optimizations for online video analysis. 
Focus \cite{hsieh2018focus} supports low-cost ingest-time analytics on live video that later facilitates low-latency queries. Liu et al. \cite{liu2019edge} design an online system that enables high-accuracy object detection at 60 FPS.
Reducto \cite{li2020reducto} designs an on-camera filtering module to enable real-time video analytics pipelines.

\noindent \textbf{Video Understanding with VLMs.}
% VLMs’ visual-linguistic reasoning has driven their broad use in video understanding. 
LLaVA-OneVision \cite{li2024llava} presents a family of open VLMs for video-based visual tasks.
AKS \cite{tang2025adaptive} employs an adaptive optimization algorithm to enhance the coverage of selected keyframes. BOLT \cite{liu2025bolt} introduces an adaptive frame selection strategy based on inverse transform sampling. Goldfish \cite{ataallah2024goldfish} design a methodology tailored for comprehending videos of arbitrary lengths.

\noindent \textbf{Retrieval-Augmented Generation for Video Understanding.}
Recent works apply RAG \cite{gao2023retrieval} to VLMs for long video understanding. 
VideoRAG \cite{luo2024video} retrieves three types of visually-aligned auxiliary texts to boost VLM performance. DrVideo \cite{ma2024drvideo} reformulates video understanding as long-document reasoning by converting videos into textual form. 
Ren et al. \cite{ren2025videorag} introduce graph-based textual grounding to capture cross-video semantic relationships for long video understanding.

\section{Conclusion} 
\texttt{Venus} employs an edge–cloud disaggregated architecture that enables real-time on-device perception, hierarchical historical memory construction, and efficient keyframe retrieval, forwarding only the most relevant content to cloud-hosted VLMs for accurate reasoning.
Our extensive evaluation demonstrates that \texttt{Venus} achieves $15\times$–$131\times$ speedup in total response latency compared to state-of-the-art methods.

\section*{Acknowledgments}
This work was supported in part by the National Science Foundation of China (No. U25B2002); Guangdong S\&T Programme (Grant No. 2024B0101040007); Guangdong Basic and Applied Basic Research Foundation (No. 2023B1515120058); Guangzhou Basic and Applied Basic Research Program (No. 2024A04J6367); The Program for Guangdong Introducing Innovative and Entrepreneurial Teams (No. 2017ZT07X355). National Natural Science Foundation of China (Grant No. 623B2093).

\bibliographystyle{IEEEtran}
\normalem
\bibliography{reference}

\end{document}